\documentclass[conference]{IEEEtran}
\IEEEoverridecommandlockouts
\usepackage{cite}
\usepackage{amsmath,amssymb,amsfonts}
\usepackage{bm}
\usepackage{nicefrac}
\usepackage{siunitx}
\usepackage{graphicx}
\usepackage{textcomp}
\usepackage{xcolor}
\usepackage{subcaption}
\usepackage[ruled,linesnumbered]{algorithm2e} % vlined, vlined
\def\BibTeX{{\rm B\kern-.05em{\sc i\kern-.025em b}\kern-.08em
    T\kern-.1667em\lower.7ex\hbox{E}\kern-.125emX}}

\usepackage{cite}
\usepackage{subcaption}
    
\usepackage{tikz}
\usetikzlibrary{math}
\usetikzlibrary{shapes.geometric,calc}
\usetikzlibrary{arrows.meta, quotes, angles}
\usetikzlibrary{decorations}
\usetikzlibrary{chains,shapes.misc,positioning,scopes}
\usepackage{circuitikz}
\usetikzlibrary{circuits.ee.IEC}
\usetikzlibrary{decorations}
\usetikzlibrary{dsp}
\usetikzlibrary{shadows}
\usepackage{subcaption}

\newcommand{\acs}{\frac{A}{C_\mathrm{s}}}
\newcommand{\nacs}{\nicefrac{A}{C_\mathrm{s}}}
\newcommand{\csa}{\frac{C_\mathrm{s}}{A}}

\definecolor{mygreen}{RGB}{80, 220, 100}

\renewcommand{\baselinestretch}{0.97}    
    
\begin{document}

% End-to-end Channel Model for Controlled-Release Drug Delivery using Practical Drug Carriers
% 
\title{Channel Modeling for Drug Carrier Matrices
%\title{On the Assessment of the Channel Response for a Spherical Drug Carrier Matrix
\thanks{This work was supported in part by the German Research Foundation (DFG) under grant number SCHO 831/14-1}
}

\author{\IEEEauthorblockN{Maximilian Sch\"{a}fer\textsuperscript{*}, Yolanda Salinas\textsuperscript{\textdagger}, Alexander Ruderer\textsuperscript{*}, Franz Enzenhofer\textsuperscript{\textdagger},\\ Oliver Br\"{u}ggemann\textsuperscript{\textdagger}, Robert Schober\textsuperscript{*}, and Werner Haselmayr\textsuperscript{\textdagger}}
\IEEEauthorblockA{\textit{\textsuperscript{*}Friedrich-Alexander University Erlangen-N\"urnberg (FAU),} Erlangen, Germany 
%		Telecommunications Laboratory (LNT), 
\\
	\textit{\textsuperscript{\textdagger}Johannes Kepler University Linz (JKU),}
	Linz, Austria}
	\vspace*{-0.8cm}
}

\maketitle

% ===================================================================

\begin{abstract}
Molecular communications is a promising framework for the design of controlled-release drug delivery systems. 
In this framework, drug carriers are modeled as transmitters, the diseased cells as absorbing receivers, and the channel between transmitter and receiver as diffusive channel. 
However, existing works on drug delivery systems consider only simple drug carrier models, which limits their practical applicability. 
In this paper, we investigate diffusion-based spherical matrix-type drug carriers, which are employed in practice. 
In a matrix carrier, the drug molecules are dispersed in the matrix and diffuse from the inner to the outer layers of the carrier once immersed in a dissolution medium. 
We derive the channel response of the matrix carrier transmitter for an absorbing receiver and validate the results through particle-based simulations. 
Moreover, we show that a transparent spherical transmitter, with the drug molecules uniformly distributed over the entire volume, is as special case of the considered matrix system. 
For this case, we provide an analytical expression for the channel response. 
Finally, we compare the channel response of the matrix transmitter with those of point and transparent spherical transmitters to reveal the necessity of considering practical models. 
\end{abstract}

%\begin{IEEEkeywords}
%do we need key words
%\end{IEEEkeywords}

% ===================================================================
\section{Introduction}
\label{sec:intro}
Molecular communications (MC) considers the transmission of information using biochemical signals over multiple scales~\cite{Nakano_13}. 
Over the past few years, the MC paradigm was exploited to gain more insight into the operation of biological systems, to control the behavior of such systems, and for the design and implementation of synthetic MC systems in the macro- and micro-/nanoscale~\cite{Farsad_16}. 
Human-made MC systems are expected to have applications in biomedical, environmental, and industrial engineering~\cite{Nakano_13, Farsad_16}. % BOOK: Chude_Okonkwo_19
%Widely found in nature  as it is essential for all living entities to retain their functionalities, for example the endocrine system releases signaling molecules (i.e., hormones) into the blood stream to communicate with distant target cells in order to regulate body activities. 

The MC framework is a promising approach to design drug delivery systems, where nanoparticle carriers deliver drug molecules to diseased cell sites and release them at the right time and rate~\cite{Chude_Okonkwo_17,Chahibi_17,Sutradhar2016,salinas_2020}. 
This reduces potential side effects on non-target sites and helps mitigate toxicity and drug wastage, compared to conventional treatment. 
Based on the MC paradigm, drug carriers are modeled as transmitters (TXs), diseased cells are modeled as receivers~(RXs), and the drug propagation between TX and RX is modeled as a random channel~\cite{Chude_Okonkwo_17,Chahibi_17}. 
The MC related research on drug delivery systems can be categorized into three areas: \textit{i) target detection} aims to develop methods for localization of diseased cells and moving the drug carriers towards them~\cite{Nakano_17}; \textit{ii) drug propagation in the circulatory system} aims to develop models for the distribution of drug molecules or  carriers over time for the optimization of drug injection~\cite{Chahibi_13}; \textit{iii) controlled local drug release} aims to design an optimum controlled-release profile, assuming that the drug carriers are already near the diseased cells~\cite{Femminella_15,Salehi_17,Salehi_18,Cao_19,Lin_21}. 
In this work, we focus on \textit{controlled local drug release}.
Although the existing works on controlled-release drug delivery systems consider various effects, such as mobility~\cite{Cao_19} and limited reservoir capacity~\cite{Salehi_17,Salehi_18}, only simple models for the drug carriers have been considered, such as point and simple spherical TXs~\cite{noel:16}. 
Hence, the objective of this work is to introduce and assess models for practical drug carriers. 
In particular, we consider a diffusion-based spherical matrix-type drug carrier~\cite{Macha_20}. In such matrix systems, the drug molecules are dispersed in a matrix, usually a polymer\footnote{Polymers are mostly used as material due to their versatile properties~\cite{Macha_20}.}. 
The matrix can be either homogeneous or porous and the release of the drug molecules is mainly influenced by the diffusion in the matrix. 
The drug molecules are dissolved from the matrix and diffuse from the inner to the outer layers of the carrier before propagating further into the surrounding medium. 
Since the diffusion distance increases for the drug molecules located further inside the carrier, the release rate is not constant and decreases over time. 
There are various models for the drug molecule release from homogeneous and porous matrix systems, see~\cite{Lee_11} for a comprehensive overview. However, these models do not consider the propagation of the released molecules into the 
surrounding environment towards a RX, which is crucial for the investigation of drug delivery systems based on the MC paradigm.

In this paper, we investigate a diffusive MC system with a homogeneous matrix system as TX and an absorbing~RX~(e.g.,~diseased cell). 
Our results provide the basis for the design of practical controlled-release drug delivery systems. 
Our main contributions can be summarized as follows:
\begin{itemize}
  \item We discuss and analyze existing models for the release of molecules from spherical homogeneous matrix systems.
  \item We derive an expression for the channel response (CR) for a matrix TX and an absorbing RX in a three-dimensional~(3D) unbounded environment. %The obtained analytical expression includes a tuning parameter that controls the rate of the drug release.
  \item For the special case where molecules are instantaneously released from the matrix TX, we provide an analytical expression for the CR. 
  This result corresponds to the CR of a transparent spherical TX, which was discussed in~\cite{noel:16}, but a solution could only be obtained by numerical integration.
  \item We develop a particle-based simulation (PBS) for the matrix TX to verify our theoretical results and compare the CR and the absorption rate of the matrix~TX with those of point and transparent spherical~TXs.
\end{itemize}

% REMOVE organization to save space
%The remainder of this paper is organized as follows: In Section~\ref{sec:drug}, we present models for the drug release in homogeneous matrix systems. Based on this model,  we derive the channel response of a matrix TX for an absorbing RX in Section~\ref{sec:channel}. Numerical evaluations of the channel response of a matrix TX and a validation through a particle-based simulation is presented in Sec.~\ref{sec:eval}. Moreover, a comparison with a point and virtual sphere TX is provided. Finally, Section~\ref{sec:concl} concludes the paper.

% ===================================================================
\section{Spherical Homogeneous Matrix System} %Drug Release of Homogeneous Matrix Systems
\label{sec:drug}
%\vspace*{-0.5ex}
In this section, we present different models for the release of molecules from spherical homogeneous matrix systems, which serve as realistic TX models for the gradual release of molecules. First, we provide a 
mathematical formulation of the molecule dissolution process from the matrix. Then, we present two promising solutions from the literature on controlled drug release~\cite{Lee_80,Frenning:2004}, which are utilized as TX models in Section~\ref{sec:channel}. 
Finally, we propose a PBS model for the gradual molecule release from matrix systems, which is used in Section~\ref{sec:eval} for the validation of the theoretical results.
%
%
%investigate the molecule release from homogeneous matrix systems. 
%Depending on the structure (homogeneous or porous) of the matrix and the initial drug loading, several approximate analytical \cite{Lee_80, Higuchi:1963} and numerical \cite{Koizumi:1995} models have been derived, see \cite{Lee_11} for an overview.  
%\textcolor{red}{Revise. Hinweis auf ''not investigated in MC so far'', and focus on the TX! }

\subsection{System Description and Preliminaries}
%\vspace*{-0.5ex}
%Fig.~\ref{fig:nanogel} shows an illustration of a polymeric nanogel loaded with drug molecules, a practical drug carrier \cite{referenz to be added}. 
%Initially, the drug molecules are dispersed in the nanogel. When the carrier is placed in a solution, the drug molecules are dissolved from the polymeric structure and diffuse inside the nanogel until they are finally released into the surrounding medium. 

\begin{figure}[t]
	\centering
	\scalebox{0.78}{
	\begin{tikzpicture}[axis/.style={thick, dspconn, >=stealth'},]
		% axis
    	
        \draw[thick, fill=blue!10!white] (0,0) circle (1.5);
        \draw[fill =black!80!white, draw = black!80!white] (1.1,0.9) circle (1pt);
        \draw[fill =black!80!white, draw = black!80!white] (0.8,0.8) circle (1pt);
        \draw[fill =black!80!white, draw = black!80!white] (1.3,1.1) circle (1pt);
        \draw[fill =black!80!white, draw = black!80!white] (1.2,1.2) circle (1pt);
 		\draw[fill =black!80!white, draw = black!80!white] (0.9,0.7) circle (1pt);       
 		\draw[fill =black!80!white, draw = black!80!white] (0.7,0.8) circle (1pt);
		\draw[fill =black!80!white, draw = black!80!white] (1.0,0.6) circle (1pt);
		\draw[fill =black!80!white, draw = black!80!white] (1.2,1) circle (1pt);
		\draw[fill =black!80!white, draw = black!80!white] (1.2,0.65) circle (1pt);

        \draw[fill = blue!60!white, very thick, dashed] (0,0) circle (1);
        \draw[axis] (-0.3,0)  -- (2,0) node(xline)[below right]
        {$x$};
        \draw[dspline] (1.5,0.1) -- (1.5,-0.1) node[below right]{$a$};
        \draw[dspline] (0,0.1) -- (0,-0.1) node[below right]{$0$};
        \node[inner sep = 0pt](r0) at (-3, -1){dissolved drugs};
        \draw[dspconn] (r0) -- (-1.2,-0.3);
        \node[inner sep = 0pt](rt) at (1.8, 2.3){undissolved drugs};
        \draw[dspconn] (rt) -- (0.2,0.6);
        
        \node[inner sep = 0pt](t0) at (-1.4, 2.3){initial position $R(0)$};
        \draw[dspconn] (t0) -- (-0.5,1.4);
        \node[inner sep = 0pt](t1) at (-2.6, 1.7){diffusing front $R(t)$};
        \draw[dspconn] (t1) -- (-0.9,0.5);
        
        \draw[dspconn, very thick] (1.2,0.7) -- (2,1.3) node[right]{drug release};
	\end{tikzpicture}
	}
	\vspace*{-0.5ex}
	\caption{\small Schematic release of a dispersed drug from a spherical non-erodible homogeneous matrix system (shown in 2D).}
	\label{fig:1}
	\vspace*{-2ex}
\end{figure}
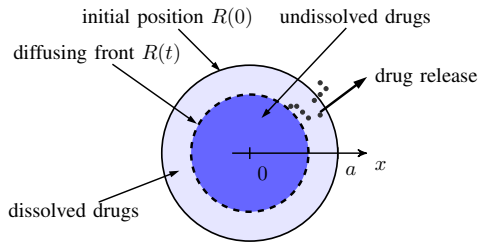

In the following, we discuss models for the release process of molecules from a \textit{spherical homogeneous non-erodible matrix system}. Fig.~\ref{fig:1} shows a two-dimensional schematic of this process. The matrix of radius $a$ in Fig.~\ref{fig:1} is initially loaded with undissolved molecules (dark blue).
%, and the initial loading is denoted by $A$. 
The release process in a solution 
%with drug solubility $C_\mathrm{s}$ 
is modeled by a diffusing front $R(t)$ that defines the time dependent dissolution of molecules in the matrix \cite{Higuchi:1963}. Dissolved molecules diffuse inside the matrix until they enter the surrounding medium at $x = a$. 
For the modeling of the release process, we make the following assumptions:
\begin{itemize}
	\item[A1)] Due to the \textit{homogeneous release} from the spherical matrix into an unbounded environment, the 3D system can be reduced to a one-dimensional system as both angular components can be neglected \cite{Lee_80}.
	\item[A2)] Since the molecule carrying matrix is \textit{non-erodible}, the impact of an additional inward moving eroding front~(see~\cite[Fig.~1]{Lee_80}) is negligible. 
	\item[A3)] The moving boundary $R(t)$ (dashed line in Fig.~\ref{fig:1}) separates undissolved dispersed molecules (dark blue) and dissolved molecules (light blue).
\end{itemize}

The molecule release rate from the matrix mainly depends on the ratio $\nacs$, with initial loading per unit volume $A$ and solubility in the surrounding medium $C_\mathrm{s}$ \cite{Higuchi:1963}. 
%For example, a
Assuming a given initial loading $A$, molecules dissolve slowly for small values of solubility $C_\mathrm{s}$ (high $\nicefrac{A}{C_\mathrm{s}}$) and fast for large values of $C_\mathrm{s}$ (small $\nicefrac{A}{C_\mathrm{s}}$). A special case is~\mbox{$A = C_\mathrm{s}$},~i.e.,~$\nacs = 1$, where all molecules are dissolved and released instantaneously when the carrier is placed into the medium. This case is equivalent to the instantaneous release of molecules from a transparent sphere, which has been discussed in \cite{noel:16}.
%
%In the following, the cases of $\nacs >1$ and \mbox{$\nacs = 1$}, corresponding to gradual and instantaneous release, are discussed separately.

\subsection{Gradual Release Process}
\label{sec:drug_cons}

In the following, we investigate models for the gradual and instantaneous release of molecules, i.e., $\nacs \geq 1$. After providing a mathematical description of the problem, we review two models from the literature on controlled drug release that are promising for their application as TX models in the MC context.

Based on assumptions A1) -- A3), the release process in Fig.~\ref{fig:1} can be described by a moving boundary problem~\cite{Higuchi:1963}. In particular, for a spherical homogeneous non-erodible matrix system with initial loading $A$ and solubility $C_\mathrm{s}$, the concentration $C(x,t)$ in the matrix can be described by the following partial differential equation~\cite[Eq.~(1)]{Lee_80}
\begin{align}
	\frac{\partial C}{\partial t} = x^{-2}\frac{\partial}{\partial x}\left(x^2 D \frac{\partial C}{\partial x}\right),
	\label{eq:1}
\end{align}
where $D$ is the diffusion coefficient of the surrounding dissolution medium. As shown in Fig.~\ref{fig:1}, $x$ is the radial coordinate, the center of the sphere is at $x = 0$ and the surface at $x = a$.
% In \eqref{eq:1}, the actually 3D problem is reduced to a 1D problem referring to assumption A1.
Assuming equilibrium between the moving diffusion front and the environment, the boundary conditions are given by~\cite{Lee_80}
\begin{align}
	C\big\vert_{x = a} &= 0,\qquad
	C\big\vert_{x = R(t)} = C_\mathrm{s}, 
	\label{eq:2}\\[1ex]
	D\frac{\partial C}{\partial x}\big\vert_{x = R(t)} &= (A - C_\mathrm{s}) \frac{\partial R}{\partial t}, \label{eq:3}
\end{align}
where $R(t)$ is the time-dependent position of the diffusion front, with initial position 	$R(0) = a$. Moreover, the first condition in~\eqref{eq:2} describes the perfect sink condition. 

An analytical solution of the moving boundary problem~\mbox{\eqref{eq:1}--\eqref{eq:3}} has not yet been reported. 
However, a numerical solution has been presented in \cite{Koizumi:1995} where the finite difference method~(FDM) is used to solve the moving boundary problem, which provides very accurate results.  
Therefore, the solution from \cite{Koizumi:1995} is used as ground truth for the numerical evaluation in Section~\ref{sec:eval}. 
However, the FDM requires high computational effort and, thus, different approximate analytical solutions have been presented in \cite{Lee_80,Higuchi:1963,Koizumi:1995, Frenning:2004}. 
All solutions have in common that their accuracy is rather poor for small~$\nacs$ values, corresponding to a fast release process. But for larger $\nacs$ values, which are typically of interest in practice, the accuracy of the approximate solutions improves~\cite{Lee_11}
%\footnote{We note that for practical drug carriers, the initial drug loading $A$ is higher than the drug solubility $C_\mathrm{s}$. Therefore, the solutions in \cite{Lee_80,Higuchi:1963,Koizumi:1995,Frenning:2004} are usefull approximations.}
.
 
A promising solution is presented in \cite{Lee_80}, as it is valid for the widest range of $\nacs$ ratios.
It is obtained by normalization of \eqref{eq:1}--\eqref{eq:3}, and the application of a \textit{double-integration heat balance integral}\footnote{The individual steps of the solution are not shown here for brevity, but are given in \cite{Lee_80}.} approach \cite{Langford:1971}. 
This results in an analytical expression for the amount of molecules $M$ that are adsorbed at the surface of the matrix~(perfect sink), which is identical to the amount of molecules released into the surrounding medium at $x = a$. 
The fraction of released molecules can be expressed as a function of the normalized diffusion front position~$\delta = 1 - \nicefrac{R}{a}$ and is given by~\cite[Eq.~(28)]{Lee_80}
\vspace*{-1ex}
\begin{align}
	\frac{M(\delta)}{M_\infty}&\bigg\vert_{\acs \geq 1} \!\!= \left[1 -(1-\delta)^3 \right]\left(1 - \csa \right)  + \nonumber\\
	&3\delta\csa\left[\left( a_1 + \frac{a_2}{2} + \frac{a_3}{3}\right) - \left(\frac{a_1}{2} + \frac{a_2}{3} +\frac{a_3}{4}\right)\delta\right]\!, \label{eq:9} 
\end{align}
where $M_\infty = A \frac{4}{3}\pi a^3$ is the total amount of available molecules and coefficients $a_1, a_2$, and $a_3$ are given by \cite[Eqs.~(19), (19a)]{Lee_80}
\vspace*{-1ex}
\begin{align}
	a_1 &= 1, &a_2 &= -a_3 -1,\label{eq:11}\\
	a_3 &= \lambda - \sqrt{\lambda^2 - 1}, 
	&\lambda &= 1-\left( 1 - \acs\right)(1-\delta) \label{eq:12}.
\end{align}
We note that the result in \eqref{eq:9} takes into account that dissolved molecules are reflected at the undissolved core of the matrix. Moreover, when the diffusion front reaches~$x = 0$ all molecules are dissolved and \eqref{eq:9} should yield one for~$\delta = 1$, but it gives $\nicefrac{M(1)}{M_\infty} = 1 - \nicefrac{C_\mathrm{s}}{(4A)}$. This confirms the previously mentioned approximate character of \eqref{eq:9} \cite{Lee_80}, where the accuracy increases with increasing~$\nacs$. 
Complementary to~\eqref{eq:9}, the normalized position of the diffusion front $\delta$ can be expressed as a  function of the time as follows \cite[Eq.~(25)]{Lee_80}
\begin{align}
	\frac{D t}{a^2} = \frac{1}{12}\left[6\frac{A}{C_\mathrm{s}} - 4 -a_3\right]\delta^2 - \frac{1}{3}\left(\acs - 1\right)\delta^3. \label{eq:10}
\end{align}
%The coefficients $a_1, a_2$, and $a_3$ in \eqref{eq:9}, \eqref{eq:10} follow 
%%from the a polynomial approximation of normalized concentration $\theta$ and can be derived 
%as \cite[Eqs.~(19), (19a)]{Lee_80}
%\begin{align}
%	a_1 &= 1, &a_2 &= -a_3 -1,\label{eq:11}\\
%	a_3 &= \lambda - \sqrt{\lambda^2 - 1}, 
%	&\lambda &= 1-\left( 1 - \acs\right)(1-\delta) \label{eq:12}.
%\end{align}
Setting $\delta = 1$ in \eqref{eq:10} yields the time duration $t_\mathrm{rel}$ needed for all molecules to be released from the matrix 
\begin{align}
t_\mathrm{rel} = \frac{a^2}{D}\left(\frac{1}{6}\acs - \frac{1}{12}\right). \label{eq:release_time}
\end{align}
%For practical application it would be desirable to have a single function that describes the drug release over time. 
To describe the amount of molecules released over time, \eqref{eq:9} and \eqref{eq:10} have to be evaluated simultaneously. 
Exploiting that the diffusion front $\delta$ ranges from $0$ to $1$, i.e., $R(0) = a$ and $R(t_\mathrm{rel}) = 0$, allows to calculate the number of released molecules $M$ at time $t$ based on \eqref{eq:9} and \eqref{eq:10}. 
We note that \eqref{eq:9} and \eqref{eq:10} are only valid for $t\in[0,t_\mathrm{rel}]$, because $t > t_\mathrm{rel}$ would correspond to a negative position of the diffusion front $R(t)$, which is physically not possible.

For the design of MC systems using the matrix system as practical TX model, it is desirable to describe the normalized number of released molecules $\nicefrac{M}{M_\infty}$ in \eqref{eq:9} as an explicit function of time. 
In \cite{Frenning:2004}, an approximate function for the  release over time has been proposed for large values of $\nacs$. Assuming~$\nacs \gg 1$ allows to simplify~\eqref{eq:10} such that it can be solved for the non-normalized diffusion front position $R(t)$ in terms of a cubic root~\cite[Sec.~2.3]{Frenning:2004}
\begin{align}
	\frac{R}{a} = &\frac{1}{2}\left(1 - \frac{1}{3}\csa\right)\nonumber\\
		&+\! \left(1 \!+\! \frac{1}{3}\csa\right) \!\cos\!\left(\frac{\arccos\left(12\frac{C_\mathrm{s}D}{A a^2}t - 1\right) + 4\pi}{3}\right)\!, \label{eq:frenning:delta}
\end{align} 
$t\in [0, t_\mathrm{rel}]$. Applying the same assumption, the normalized number of released molecules \eqref{eq:9} can be simplified as \cite{Frenning:2004}
\begin{align}
	\frac{M(t)}{M_\infty}\bigg\vert_{\acs \gg 1} \!\!\!\!\!\!= 1 \!-\! \left(\frac{R}{a}\right)^3 \!\!\!+ \frac{1}{2}\frac{C_\mathrm{s}}{A}\!\left[2\left(\frac{R}{a}\right)^3 \!\!\!-\! \left(\frac{R}{a}\right)^2 \!\!\!-\! \frac{R}{a} \right]\!\!, \label{eq:frenning:M}
\end{align}
$t\in [0, t_\mathrm{rel}]$. An explicit closed-form expression for the normalized number of released molecules as a function of time $t$ can be obtained by inserting \eqref{eq:frenning:delta} into \eqref{eq:frenning:M}. 
The validity of the simplified model \eqref{eq:frenning:delta}, \eqref{eq:frenning:M} is further discussed in Section~\ref{sec:eval}.
%Similar to \eqref{eq:9}, is \eqref{eq:frenning:M} only valid for $t\in [0, t_\mathrm{rel}]$.
%It is important to note, that \eqref{eq:frenning:delta} is only valid until all molecules are dissolved from the matrix, i.e., $\nicefrac{M}{M_\infty}= 1$. Therefore, \eqref{eq:frenning:M} can only be evaluated for $t\in [0, t_\mathrm{rel}]$.   

\subsection{Instantaneous Release Process}
\label{sec:drug_ins}

Next, we consider the special case $\nacs = 1$, where the molecule solubility $C_\mathrm{s}$ is equal to the initial loading $A$. 
This means the diffusion front immediately reaches the center of the matrix. 
This is equivalent to the case, where molecules are distributed uniformly over the entire volume of a transparent sphere and are then instantaneously released.
Therefore, the release process is no longer a moving boundary problem, and simpler mathematical descriptions can be used instead.

The number of molecules released from a sphere, uniformly filled with molecules, into a bounded release medium is given in~\cite[Eq.~(6.30)]{crank:1975}. This solution can be extended to an unbounded release medium, which corresponds to the normalized number of released molecules of a matrix for the special case $\mbox{$\nacs = 1$}$~\cite{Frenning:2004} 
\begin{align}
	\frac{M(t)}{M_\infty}\bigg\vert_{\acs = 1} = 1 - \frac{6}{\pi^2}\sum\limits_{n=1}^\infty \frac{1}{n^2}\mathrm{exp}\left(\gamma_n t\right),
	 \label{eq:crankVinf1}
\end{align}
where $\gamma_n = -Dn^2\frac{\pi^2}{a^2}$. 

\begin{algorithm}[t!]
\DontPrintSemicolon
% Determine diffusion front $R(t_k)$ using FDM \cite{Koizumi:1995} \\
 Distribute molecules uniformly in the spherical TX \;
 Determine diffusion front $R(t)$ using FDM~\cite{Koizumi:1995} \;
 \For{$k \leftarrow$ 1 \KwTo $K$ }{ 
  \For{$m \leftarrow$ 1 \KwTo $M_\infty$}{
   \If{molecule is not marked as released}{
     \If{$d_m(t_k) \geq R(t_k)$}{
     $\mathbf{r}_m(t_{k+1}) =\mathbf{r}_m(t_k) + \mathcal{N}(\mathbf{0},2D\Delta t \mathbf{I})$ \;
       \If{$d_m(t_k) < R(t_k)$}{
       $\mathbf{r}_m(t_{k+1}) = \mathbf{r}_m(t_k)$\tcp*{\footnotesize reflection}
       }
     }
     } 
     \If{$d_m(t_k) \geq a$}{
     $M(t_k) = M(t_k) + 1$ \;
     Mark molecule as released
     }
   }
 }
 \caption{PBS to determine the number of molecules released from matrix systems}
%\caption{PBS for molecule release from matrix systems}
 \label{alg:pbs}
\end{algorithm}
% ----- HACK-ME -----

\subsection{Particle-Based Simulation}
\label{sec:drug_pbs}
To validate the expressions for the number of molecules released from matrix systems presented in \eqref{eq:9}, \eqref{eq:frenning:M}, and~\eqref{eq:crankVinf1}, we have developed a PBS model. The simulator was implemented 
    in the programming language Python and time-consuming parts were realized using Cython. The individual simulation steps are summarized in Algorithm~\ref{alg:pbs}. 
    The simulation accounts for reflections at the undissolved matrix core, assuming that the molecule bounces back to its previous position~\cite{Deng_16}. This was also taken into account for the derivation of the theoretical results presented above. 
After the simulation, the results from multiple simulation runs are accumulated and averaged.
%    The simulation algorithm is executed in parallel and after the simulation, the results from different simulation runs are accumulated and averaged.

To keep the presentation of Algorithm~\ref{alg:pbs} concise we use the following definitions: The time step 
 is denoted by~$\Delta t$ and the discrete time instances are defined by $t_k=k\Delta t$. Moreover, the maximum number of time steps is given by~$K$. The position of the $m$th molecule at time $t_k$ is defined by~$\mathbf{r}_m(t_k)$ and its Euclidean distance to the origin of the sphere~$\mathbf{r}_\text{s}$ is calculated as $d_m(t_k) = ||\mathbf{r}_m(t_k) - \mathbf{r}_\text{s}||_2$. 
 The molecule movement is modeled as random walk, where the molecule position is updated by~$\mathcal{N}(\mathbf{0},2D\Delta t \mathbf{I})$. Here, $\mathcal{N}(\pmb{\mu}, \pmb{\Sigma})$ denotes a multivariate Gaussian distribution with mean vector $\pmb{\mu}$ and covariance matrix $\pmb{\Sigma}$, and $\mathbf{0}$ and $\mathbf{I}$ denote the all-zero vector and the identity matrix, respectively. 

%The simulation algorithm initially distributes the molecules uniformly over the entire volume (i.e., all molecules are undissolved). Then, the molecules are gradually released, with the individual simulation steps summarized in Algorithm~\ref{alg:pbs}. In particular, the procedure for each molecule at each time step $\Delta t$ is shown. The diffusion front position $R(t)$ is determined in advance at times~$t_k=k\Delta t$ using the FDM~\cite{Koizumi:1995}. Similar to the presented theoretical expression, the PBS considers reflections at the undissolved matrix core by undoing the last movement~\cite{Deng_16}. 

% ===================================================================
\section{Channel Response for Spherical Matrix}
\label{sec:channel}

\begin{figure}[!t]
	\vspace*{1ex}
	\centering
	\includegraphics{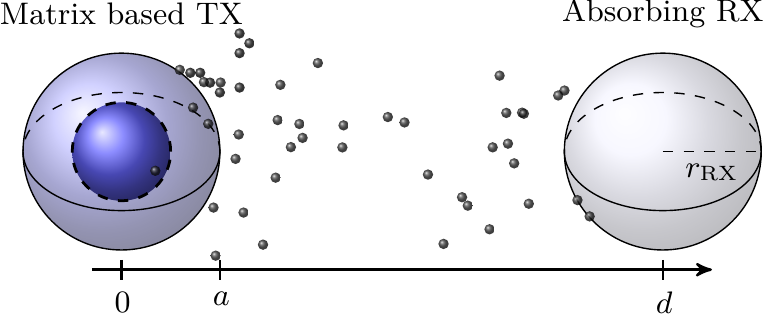}
	\vspace*{-1ex}
	\caption{\small System model for the molecule release from a spherical matrix in a 3D unbounded environment with an absorbing RX.}
	\label{fig:system}
%	\vspace*{-2ex}
\end{figure}

In Section~\ref{sec:drug}, the gradual and instantaneous release of molecules from a spherical matrix have been analyized, assuming a perfect sink condition at the matrix boundary~(see~\eqref{eq:2}). 
In the following, we extend this study and derive the CR of a spherical matrix TX and an absorbing RX in an unbounded~3D environment, which is illustrated in Fig.~\ref{fig:system}. 

The models presented in Sections~\ref{sec:drug_cons}~and~\ref{sec:drug_ins} characterize the normalized number of released molecules from the matrix surface at $x = a$. 
Therefore, the~CR of the matrix TX can be derived by combining the molecule release of the matrix system and the CR for the instantaneous release of molecules from the surface of a transparent spherical TX. 
For the following analysis, we assume that, once molecules are released into the surrounding medium, the matrix does not impede the diffusion of the molecules, i.e., reflections at the TX are not considered\footnote{Even if a reflective TX model would be more realistic, a transparent TX surface is an appropriate approximation, see \cite{noel:16}.}.
 
\subsection{Channel Impulse Response - Surface Transmitter}

Assuming instantaneous release of the molecules from the surface of a sphere at $t = 0$, the hitting probability of the molecules at an absorbing RX of radius $r_\mathrm{RX}$ is given by \cite[Eq.~(9)]{Huang2020}
\vspace*{-1em}
\begin{align}
p_{\mathrm{s}}(t) 
&=
\frac{2\rho a r_{\mathrm{RX}}}{d}
\sqrt{\frac{\pi D}{t}}
\bigg[
\mathrm{exp}{\left(\!-\frac{\beta_1}{t} \right)} \!-\! 
\mathrm{exp}{\left(\!-\frac{\beta_2}{t} \right)} 
\bigg]\!,
\label{TS_TX_Abs_RX,Rate}
%\\
%&= f \sqrt{\frac{1}{t}} 
%\bigg[\mathrm{exp}\bigg(-\frac{g}{t}\bigg) - 
%\mathrm{exp}\bigg(-\frac{h}{t}\bigg) \bigg]
\end{align}
where $\beta_1 =\frac{ (a+r_{\mathrm{RX}})(a+r_{\mathrm{RX}}-2d) + d^2 }{4D}$ and $\beta_2 = \frac{ (a-r_{\mathrm{RX}})(a-r_{\mathrm{RX}}+2d) + d^2 }{4D}$. Here, the distance between TX and RX is denoted by $d$ (see Fig.~\ref{fig:system}), and $\rho = \left(4 \pi a ^2\right)^{-1}$. 
%We note that \eqref{TS_TX_Abs_RX,Rate} slightly differs from \cite[Eq.~(9)]{Huang2020}, because we do not consider the effect of membrane fusion, i.e., $k_\mathrm{f} = 0$ in \cite[Eq.~(9)]{Huang2020}. 
A proof for \eqref{TS_TX_Abs_RX,Rate} is given in \cite[Appendix~B]{Huang2020}.

\begin{figure*}[!t]
% ensure that we have normalsize text
\normalsize
% Store the current equation number. \setcounter{MYtempeqncnt}{\value{equation}}
% Set the equation number to one less than the one
% desired for the first equation here.
% The value here will have to changed if equations
% are added or removed prior to the place these
% equations are referenced in the main text.
\setcounter{equation}{15}
\begin{align}
N(t)\big\vert_{\acs=1}  &=
\frac{M_\infty  r_{\mathrm{RX}}}{\pi a  d} \sqrt{\pi D} \bigg[
\sqrt{t} \bigg( \mathrm{exp}\bigg(\!\!-\!\frac{\beta_1}{t} \bigg) \!-\!
\mathrm{exp}\bigg(\!\!-\!\frac{\beta_2}{t} \bigg) \bigg)
+
\sqrt{\pi}
\bigg(\sqrt{\beta_1}\mathrm{erf}\bigg(\!\sqrt{\frac{\beta_1}{t}} \bigg) \!-\!
\sqrt{\beta_2}\mathrm{erf}\bigg(\!\sqrt{\frac{\beta_2}{t}} \bigg)
+ \sqrt{\beta_2} \!-\! \sqrt{\beta_1}
\bigg)
\bigg]
\notag
\\
&+
\sum _{n=1}^\infty
\frac{3{M_\infty} r_{\mathrm{RX}} \sqrt{D}}
{2d a n^2 \pi^2 \sqrt{\gamma_n}} 
%\mathrm{exp}\left(\gamma_n t\right) 
\bigg\{ 
\mathrm{exp}\bigg(\!\gamma_n t-\!2 \sqrt{\gamma_n\beta_1} \bigg)
\bigg[
\mathrm{exp}\bigg(4 \sqrt{\gamma_n\beta_1} \bigg)
\cdot
\mathrm{erfc}\bigg(\sqrt{\gamma_n t} \!+\! \sqrt{\frac{\beta_1}{t}} \bigg) +
\mathrm{erfc}\bigg(\sqrt{\frac{\beta_1}{t}} \!-\! \sqrt{\gamma_n t}  \bigg) 
\bigg]
\notag
\\
&-
\mathrm{exp}\bigg(\!\gamma_n t- \!2 \sqrt{\gamma_n\beta_2} \bigg)
\bigg[
\mathrm{exp}\bigg(4 \sqrt{\gamma_n\beta_2} \bigg)
\cdot
\mathrm{erfc}\bigg(\sqrt{\gamma_n t} \!+\! \sqrt{\frac{\beta_2}{t}} \bigg) +
\mathrm{erfc}\bigg( \sqrt{\frac{\beta_2}{t}} \!-\! \sqrt{\gamma_n t}  \bigg) 
\bigg]
\bigg\}
\label{eq:response_ins2}
\end{align}
% Restore the current equation number. 
\setcounter{equation}{12}
% The IEEE uses as a separator
\hrulefill
% The spacer can be tweaked to stop underfull vboxes
\vspace*{-2.5ex}
\end{figure*}

\begin{figure*}
	\centering
	\begin{subfigure}[b]{0.49\linewidth}
            \centering
            	\includegraphics[width=\linewidth]{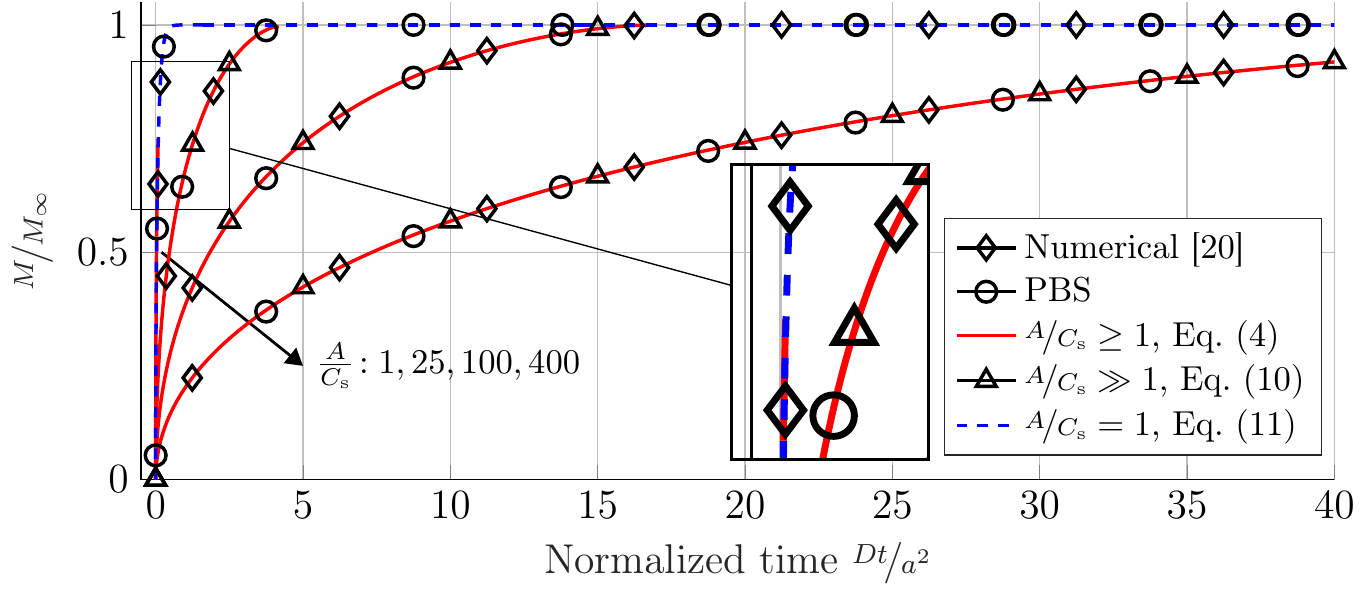}
            	\vspace*{-3.9ex}
            \caption{Normalized molecule release}
    \label{fig:profile:m}
    \end{subfigure}
    \begin{subfigure}[b]{0.49\linewidth}
            \centering
            	\includegraphics[width=\linewidth]{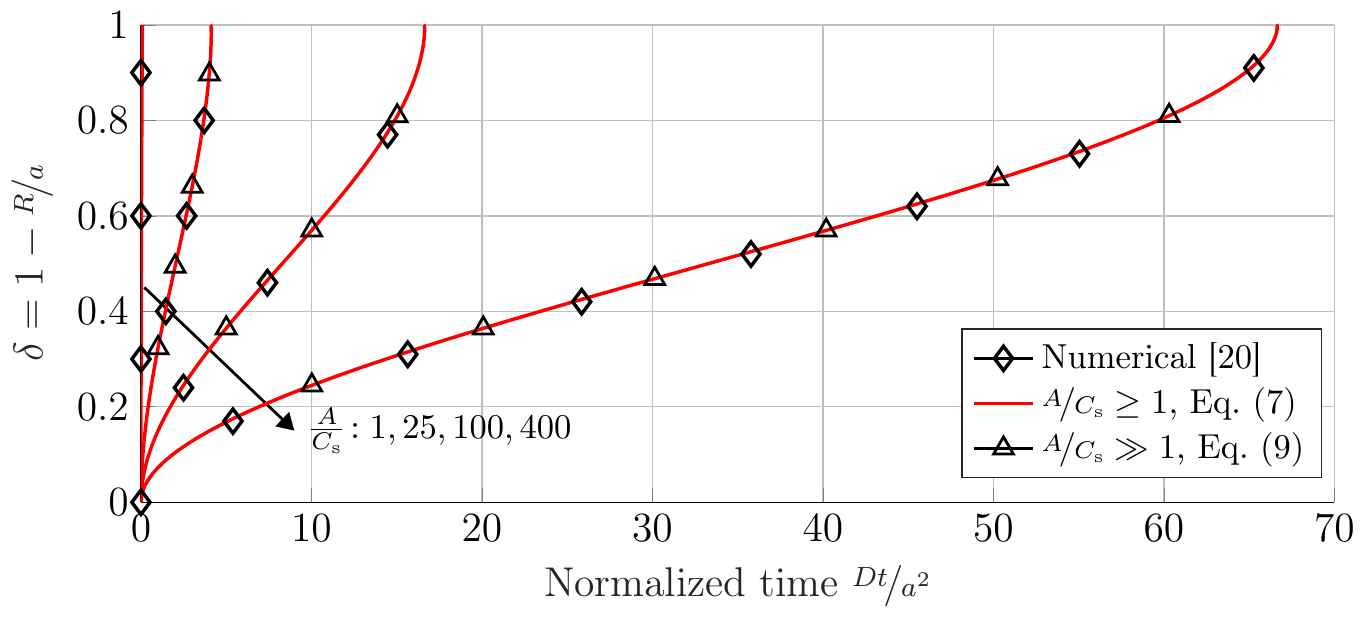}
            	\vspace*{-3.9ex}
            \caption{Diffusion front position}
    \label{fig:profile:del}
    \end{subfigure}
    \vspace*{-0.5ex}
    \caption{\small (a) Normalized release $M(t)/M_\infty$ over normalized time $\nicefrac{Dt}{a^2}$, and (b) corresponding time dependent position of diffusion front $\delta$. The FDM solution according to \cite{Koizumi:1995}, the results from PBS according to Section~\ref{sec:drug_pbs}, the approximate solutions in \eqref{eq:9}, \eqref{eq:10} and \eqref{eq:frenning:delta}, \eqref{eq:frenning:M}, and the solution for $\nacs = 1$ in \eqref{eq:crankVinf1} are considered.}
    \label{fig:profile}
     \vspace*{-3.5ex}
\end{figure*}

\subsection{Gradual Release Process}
%\vspace*{-0.5ex}
To obtain the CR for the gradual release, i.e., $\nacs > 1$, of molecules from a matrix system, we have to extend the approximate solution in \eqref{eq:frenning:M} to the interval $t\in [0,\infty)$. Exploiting that the normalized number of released molecules $\nicefrac{M}{M_\infty}$ should remain~$1$ when~$t \geq t_\mathrm{rel}$, we can express the amount of molecules released from the matrix surface as
\vspace*{-0.5ex}
\begin{align}
	\bar{M}(t)\big\vert_{\acs\gg1} \!=\! M(t)\big\vert_{\acs\gg1}\!\left(\epsilon(t) - \epsilon(t - t_\mathrm{rel})\right) + M_\infty \epsilon(t - t_\mathrm{rel}),
	\label{eq:mcon}
\end{align}
where $\epsilon(t)$ denotes the unit step function. To obtain the number of molecules absorbed at the RX, \eqref{eq:mcon} has to be convolved with the hitting probability \eqref{TS_TX_Abs_RX,Rate}, i.e.,
\vspace*{-0.8ex}
\begin{align}
	N(t)\big\vert_{\acs\gg1} = \int_0^t p_\mathrm{s}(t - \xi)\bar{M}(\xi)\big\vert_{\acs\gg1} \,\mathrm{d}\xi.
	\label{eq:response_con}
\end{align}
Due to the complex structure of $M(t)$ in \eqref{eq:frenning:M}, it is not possible to find a closed-form solution for~\eqref{eq:response_con}. Hence, to study the influence of a gradual release of molecules from the matrix system on the CR,~\eqref{eq:response_con} is evaluated numerically in Section~\ref{sec:eval}.

\subsection{Instantaneous Release Process}
%\vspace*{-0.3ex}
To obtain the CR for an instantaneous release from the matrix TX ($\nacs = 1$), \eqref{eq:crankVinf1} has to be convolved with the hitting probability in \eqref{TS_TX_Abs_RX,Rate}, i.e., 
\vspace*{-0.8ex}
\begin{align}
	N(t)\big\vert_{\acs=1} = \int_0^t p_\mathrm{s}(t - \xi)M(\xi)\big\vert_{\acs=1} \,\mathrm{d}\xi. 
	\label{eq:response_ins1}
\end{align}
The evaluation of the integral leads to an analytical expression for $N(t)$ that is given in \eqref{eq:response_ins2} where $\mathrm{erf}(x)$ and $\mathrm{erfc}(x)$ denote the error function and complementary error function, respectively.
Eq.~\eqref{eq:response_ins2}, shown at the top of the next page, specifies the number of absorbed molecules in response to an instantaneous release from a spherical matrix with $\nacs = 1$. 
We note that \eqref{eq:response_ins1} corresponds to the CR for an instantaneous release from a transparent spherical TX that has been discussed in~\cite{noel:16}, based on a point TX model.  
However, in~\cite{noel:16}, a solution for $N(t)$ could only be obtained by numerical integration.
%%%
%This is equivalent to the instantaneous release from a transparent sphere, which was considered in \cite{noel:16}, but no analytical expression for $N(t)$ was provided. 

%%% 

% ===================================================================
\section{Numerical Results}
\label{sec:eval}

%\vspace*{-0.5ex}
In the following, we numerically investigate the influence of the considered practical matrix TX model on the CR of a diffusive MC system with an absorbing RX (see Fig.~\ref{fig:system}). 
Moreover, we compare the CR of the matrix TX with those of a point TX and a transparent spherical TX~\cite{noel:16}. 
For the validation of the CR expressions for the matrix TX, we embedded the proposed PBS model of the matrix system~(see~ Section~\ref{sec:drug_pbs}) in the PBS model of a diffusive MC system with an absorbing RX \cite{noel:16}. 
All results from PBS are obtained with a time step of~$\Delta t = 1\cdot 10^{-6}\si{\second}$ and averaged over $100$ realizations.

%In this section, we numerically evaluate the CR for homogeneous matrix systems derived in Section~\ref{sec:channel} and compare it with PBS and the release from a point TX and a transparent spherical TX~\cite{noel:16}.

For the evaluation, we adopt the parameters used in~\cite{noel:16}: We consider TX and RX with radii $a=r_\mathrm{RX}=1\si{\micro\meter}$ and distances $d = \{2,\, 5\}\si{\micro\meter}$. 
The diffusion coefficient of the released molecules is 
 $D = 10^{-9}\si{\square\meter\per\second}$. 
 Initially the different TX types are loaded with $N = 10^4$ molecules, i.e., the molecules are uniformly distributed over the entire volume or are concentrated in a point. 
 At $t=0$, the molecules are instantaneously released from the point TX and the spherical TX and gradually released from the matrix TX. 
% The number of loaded molecules is $N = 10^4$. 
 Moreover, we consider the following ratios of initial loading and drug solubility $\nacs = \{1,\,25,\,100,\,400\}$.
% , with $A=\nicefrac{N}{V_\mathrm{TX}}$.
%  and $V_\mathrm{TX} = \nicefrac{4}{3}\pi r_\mathrm{TX}^3$. 

\subsection{Molecule Release Process}

First, we compare the models for the release of molecules from the matrix as described in Section~\ref{sec:drug}. 
In Fig.~\ref{fig:profile:m}, the normalized number of released molecules $M/M_\infty$ is plotted over normalized time $\nicefrac{Dt}{a^2}$ for different $\nacs$ values. 
We observe that the amount of molecules released until a certain time decreases with increasing $\nacs$, i.e., the release rate reduces with increasing~$\nacs$. 
Fig.~\ref{fig:profile:del} shows the corresponding time dependent position of the diffusion front $\delta$ derived from \eqref{eq:10} and \eqref{eq:frenning:delta}, respectively, and reveals that the speed of the diffusion front decreases with increasing $\nacs$.
From both plots in Fig.~\ref{fig:profile}, we observe that the exact FDM solution from \cite{Koizumi:1995} (diamond markers), the results from the proposed PBS according to Section~\ref{sec:drug_pbs} (circle markers), and the approximate solutions \eqref{eq:9}, \eqref{eq:10} (red curves) and \eqref{eq:frenning:M}, \eqref{eq:frenning:delta} (triangle markers) are in excellent agreement for all $\nacs > 1$. 
%Also the simplified model in \eqref{eq:frenning:M} that is a direct function of time is in excellent agreement with the exact numerical solution. 

However, for $\nacs = 1$, we observe that the approximate solution \eqref{eq:9} is not applicable as can be seen from the zoomed area in Fig.~\ref{fig:profile:m}. 
In this case, the approximate solution does not reach $1$, i.e., $\frac{M(1)}{M_\infty} = 1 - \frac{C_\mathrm{s}}{4A} = 0.75$, and is not able to capture the complete dynamics of the release process. 
Nevertheless, as $\nacs = 1$ corresponds to an instantaneous release from a transparent sphere, \eqref{eq:crankVinf1} can be applied instead of \eqref{eq:9} and is in excellent agreement with the numerical results, as shown in Fig.~\ref{fig:profile:m} (dashed blue).

\subsection{Channel Response}

\begin{figure}[t]
            \centering
            	\includegraphics[width=\linewidth]{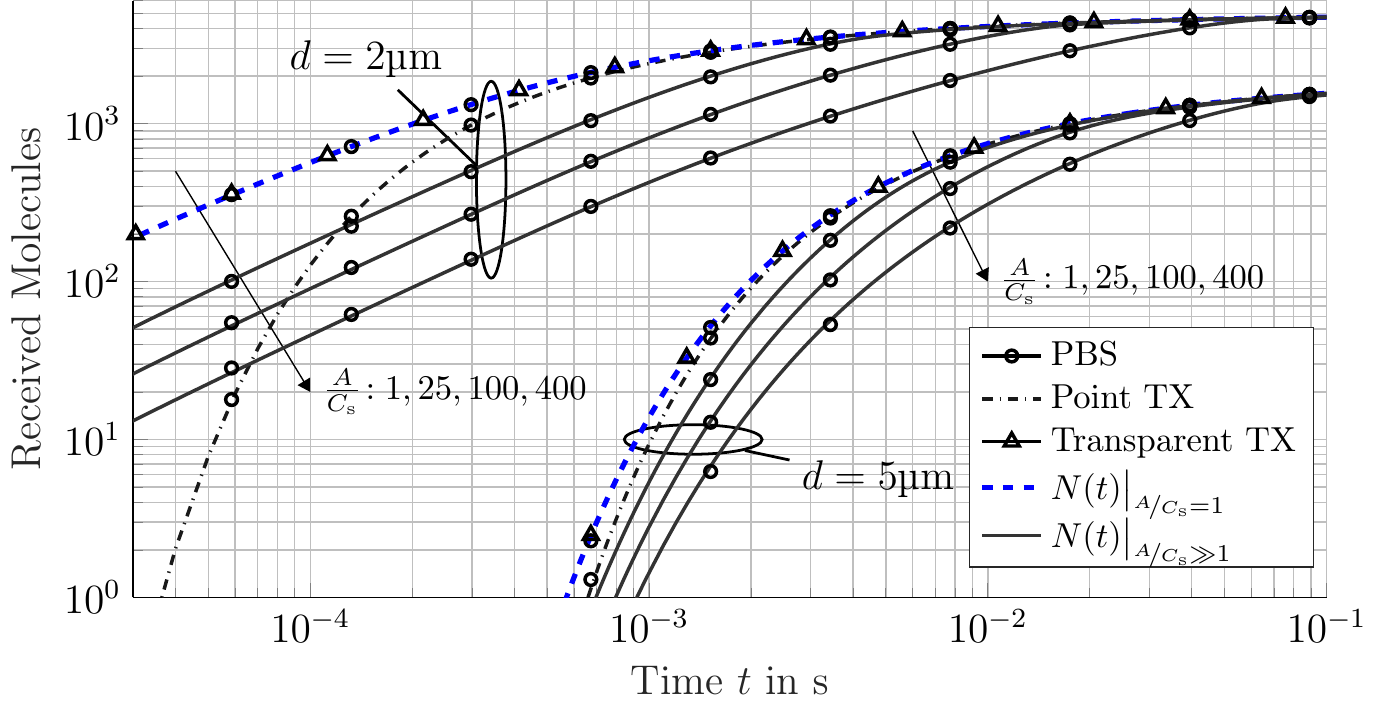} 
            	\vspace*{-3.5ex} 
 	\caption{\small Channel responses for an absorbing RX due to a point release (dash dotted), a spherical release (triangle markers), and matrix releases for different $\nicefrac{A}{C_\mathrm{s}}$-values (blue and black lines) for $d \in \{2,5\}\si{\micro\meter}$. Results from PBS are shown as circle markers.}
 	\label{fig:3}
 	 \vspace*{-3ex}
 \end{figure}

In Fig.~\ref{fig:3}, we investigate the CR of a matrix TX  for an absorbing RX as derived in Section~\ref{sec:channel}. 
The figure shows the number of molecules received by an absorbing RX for $d = \{2,\,5\}\si{\micro\meter}$ and different types of TXs, i.e., a point TX (dash dotted curves), a transparent sphere TX (triangle markers)~\cite{noel:16}, and a matrix TX with $\nacs = 1$ (Eq.~\eqref{eq:response_ins2}, dashed blue) and $\nacs \gg 1$ (Eq.~\eqref{eq:response_con}, black curves). The results from PBS are shown by circle markers for validation.

First, we observe that the results for the point, transparent, and matrix TX are in excellent agreement with the results from PBS for all considered scenarios. 
Furthermore, the differences between the considered TX types are more pronounced when the distance between TX and RX is smaller. 
The behavior of all TX types becomes similar as time increases.
For $\nacs = 1$, we observe that the CR \eqref{eq:response_ins2} is equivalent to the release from a transparent sphere. 
This confirms that \eqref{eq:response_ins2} is an analytical description for the spherical TX model proposed in \cite{noel:16}.

Moreover, we observe that for the point TX, it takes longer for the first molecules to arrive at the RX compared to the spherical and matrix TXs. 
This is because all molecules are initially located at the center of the TX and not in the outer layers of the sphere. 

For the matrix TX, we also observe that the absorption of molecules at the RX slows down for increasing $\nacs$ values.
The reason for this behavior is that for higher $\nacs$-values, the molecules dissolve more slowly from the matrix (see Fig.~\ref{fig:profile:m}). 
Moreover, for increasing $\nacs$, we observe that the delivery of molecules by the matrix TX is more spread over time compared to the point and spherical TX. 
%
%the matrix TX takes longer to deliver the molecules to the RX . 
%
%for increasing $\nacs$
%
%A similar effect can be observed for the matrix TX, whereas the effect increases as  $\nacs$ increases.   
This behavior is desirable for drug delivery systems where the amount of drugs absorbed by the RX (e.g., the diseased cell) should stay in a desired range during the delivery process \cite[Fig.~1]{Sutradhar2016}. 

\begin{figure}[t]
			\vspace*{0.3ex}
            \centering
            	\includegraphics[width=\linewidth]{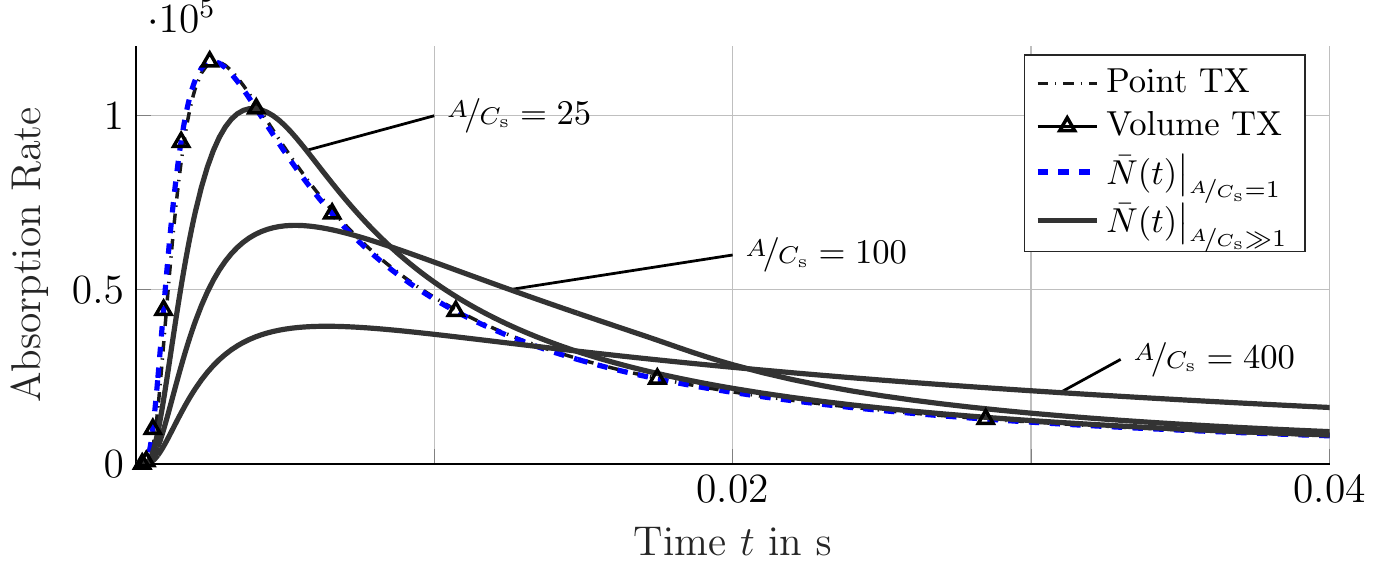}  
            	\vspace*{-3.5ex}
 	\caption{\small Absorption rate at the RX due to a point release (dash dotted), a spherical release (triangle markers), and a matrix release for different $\nicefrac{A}{C_\mathrm{s}}$ (blue and black lines) for $d = 5\si{\micro\meter}$.}
 	\label{fig:absRate}
 	 \vspace*{-3ex}
 \end{figure}

We investigate this behavior more in detail in Fig.~\ref{fig:absRate}, which shows the absorption rate\footnote{The absorption rate is defined as the time derivative of the number of received molecules, i.e., $\bar{N} = \nicefrac{\partial}{\partial t} N$.} of the RX for different TX types and $d = 5\si{\micro\meter}$. 
We observe that the absorption rates for the point, transparent, and matrix TX ($\nacs = 1$) are nearly identical. 
In particular, the absorption rate is high for small $t$, but decreases very fast as the molecule release at the TX  was instantaneous. 
For the gradual release of molecules by the matrix TX ($\nacs > 1$), we observe that the absorption rate is spread over time, because the release time \eqref{eq:release_time} of the matrix TX scales with $\nacs$. 
This reveals the importance of parameter $\nacs$ for the design of drug delivery systems with practical carriers.
In particular, the amount of delivered drugs over time can be controlled by $\nacs$, i.e., by loading $A$ and drug solubility $C_\mathrm{s}$. 
%
% i.e., by a specific variation of loading $A$ or solubility $C_\mathrm{s}$ the amount of delivered drugs over time can be controlled.

% ----------------------------------
% ----------------------------------

%\begin{itemize}
%	\item Describe considered scenario, together with a new picture (TX, RX, Channel, etc.) 
%	\item Describe simulation scenario 
%	\item Describe PBS, especially how the moving boundary is included (mention reflection at undissolved Core) 
%	\item Mention comparison to \cite{noel:16}
%	\item I suggest to use distances 2,5,10, 20 $\rightarrow$ Fig. 3 + another figure. 
%	\item Should we vary anything else in the setting? 
%	\item I suggest to show the influence on the absorption rate over time (relate to applications) 
%	\item Should we show the deviation from volume release? 
%\end{itemize}
%
%\textcolor{red}{From Here -- OLD STUFF}

% ===================================================================
\section{Conclusions}
\label{sec:concl}
In this paper, we have modeled a practical polymer-based drug carrier by a spherical homogeneous matrix system. We discussed the gradual molecule release from the matrix, which is based on a moving boundary separating dissolved and undissolved molecules. We derived expressions for the channel response of a matrix TX for an absorbing RX in an unbounded diffusive environment. 
Moreover, we derived an analytical expression for the channel response for the special case of an instantaneous release from the matrix, which is equivalent to the channel response of the well-known transparent spherical~TX. 
Our numerical evaluations showed that the channel response of the matrix TX is significantly different from that of the point TX and the transparent spherical TX. 
In particular, matrix TXs spread the release and therefore also the absorption of molecules over time. 
This reveals the necessity to take practical drug carrier models into account for the design of controlled-release delivery systems. 
All presented results have been validated by particle-based simulations. 
Interesting topics for future work include considering a reflective matrix TX surface and mobility of the TX.

\bibliographystyle{IEEEtran}
{\footnotesize
	\bibliography{./../../bib/IEEEabrv,./msnp}}

\end{document}